# Robotic fabrication of high-quality lamellae for aberration-corrected transmission electron microscopy

Hideyo Tsurusawa, [1]*† Nobuto Nakanishi,[1]† Kayoko Kawano,[1]

Yiqiang Chen,[2] Brandon Van Leer,[3] Teruyasu Mizoguchi[4]*

[1]Thermo Fisher Scientific, FEI Japan Ltd., 4-12-2, Higashi-Shinagawa, Shinagawa-ku,

Tokyo 140-0002, Japan.

[2]Thermo Fisher Scientific, Achtseweg Noord 5, 5651 GG Eindhoven, The Netherlands.

[3]Thermo Fisher Scientific, 5350 NE Dawson Creek Drive Hillsboro,OR 97124, USA.

[4]Institute of Industrial Science, University of Tokyo, 4-6-1 Komaba, Meguro-ku, Tokyo 153-8505, Japan.

†These authors contributed equally to this work.

*Correspondence should be addressed to H.T. and T.M.

E-mail: Hideyo.Tsurusawa@thermofisher.com, teru@iis.u-tokyo.ac.jp

May 24, 2021

**Abstract**

Aberration-corrected scanning transmission electron microscopy (STEM) is widely used for atomic-level imaging of materials. To accelerate the discovery of new materials based on atomic-level investigations, the throughput of aberration-corrected STEM experiments becomes more and more important. However, the throughput of the full workflow of aberration-corrected STEM is still quite low. A fundamental problem is that the preparation of high-quality thin STEM samples (lamellae) depends on manual operation. Here, inspired by the recent successes of "robot scientists", we demonstrate robotic fabrication of high-quality lamellae by focused-ion-beam (FIB) with full automation software. First, we show that robotic FIB can prepare lamellae with a high success rate, where the robotic FIB controls rough-milling, lift-out, and final-thinning processes. Then, we optimize the FIB parameters of the final-thinning process for single crystal Si. Aberration-corrected STEM imaging of these Si lamellae shows atomic-level images with 55 pm resolution. We also demonstrate robotic fabrication of high-quality lamellae of $SrTiO_3$ and sapphire. The robotic FIB system will resolve the current bottleneck of the full workflow of aberration-corrected STEM analysis and accelerate materials discovery based on atomic-level imaging.

**Introduction**

Aberration-corrected scanning transmission electron microscopy (STEM) is widely used for atomic-level imaging of materials [1, 2], offering spatial resolution in a sub-50-pm range [3, 4] and powerful micro-analysis technology [5-8]. The linkage between atomic-level structures and material properties plays a key role in materials discovery [9, 10]. To accelerate the discovery of new materials based on atomic-level structural designs, the throughput of aberration-corrected STEM experiments becomes more and more important. However, the throughput of the full workflow of aberration-corrected STEM analysis is still quite low when compared with the characterization techniques used in high-throughput experiments [11]. The low throughput of STEM experiments is mainly due to limitations in the sample preparation process. Aberration-corrected STEM usually requires a thin sample (lamella) with thickness below 50 nm and a damage-free surface. Current methods to prepare such high-quality lamellae from bulk material largely depend on manual operation and the knowledge and experience of a skilled user. So, automation of fabricating high-quality lamellae is an important challenge to accelerate the discovery of materials based on atomic-level investigations.

Focused-ion-beam (FIB) is a common method of preparing high-quality lamellae for aberration-corrected STEM imaging [12–14]. In a standard experiment, such a lamella is prepared by three main steps: rough-milling, lift-out, and final-thinning (see also Fig. 1). In the rough-milling process, FIB is used to extract a thin chunk from the bulk material. Then, the chunk is transferred to a TEM grid using an in-situ manipulator (the lift-out process). In the final-thinning process, FIB thins and polishes the chunk down to the order of 10 nm in thickness. A number of studies on functional materials have used FIB for preparing lamellae that yield high-quality atomic-level images [15–22]. FIB also allows lamella preparation from highly specific sites on the bulk sample. This is especially beneficial when the bulk sample is synthesized by a combinatorial method since the "composition" of the material depends on the "position" on the bulk sample [11, 23-26].

   An ideal solution for the high-throughput fabrication of high-quality STEM lamellae is to achieve a fully "robotic" operation of the FIB. Our concept is inspired by recent demonstrations of "robot scientists" that automatically conduct pre-determined experiments faster and more accurately than a human scientist can [27–31]. A significant challenge in robotic FIB operation is the control of the final-thinning process and optimization of the FIB parameters. Aberration-corrected STEM usually requires that the lamella should have a thickness below 50 nm and still maintain intact (damage-free) surfaces. To improve the surface quality, the FIB acceleration voltage should be reduced from 30 kV to 2 kV (or below) for the final-thinning process [12-14]. However, low-kV FIB images are typically blurred and of poor quality. This complicates the positioning of the mill (or polish) area in the FIB image and worsens the reproducibility of sample thickness and surface quality. Thus, the final-thinning process for high-quality lamellae usually depends on a careful operation by an expert. Despite pioneering efforts of semi-automatic FIB [32], fully robotic FIB operation for preparing high-quality lamellae has not been reported.

   Here, we demonstrate fully robotic FIB fabrication of high-quality lamellae. We used an FIB instrument with the latest automation software, which includes full automation of the lift-out process and improved control of low-kV FIB in the final-thinning process. We show that robotic FIB is now able to prepare lamellae from bulk materials with a high success rate. The FIB parameters of robotic operation were optimized for single crystal silicon. Aberration-corrected STEM imaging was performed on the optimized Si lamellae, showing atomic-level images with 55 pm resolution. Our results show that the resolution and sample thickness are reproducible. We also demonstrate that the robotic FIB can prepare high-quality lamellae of $SrTiO_3$ and sapphire. These results suggest that the robotic FIB system may be applicable for a wide range of materials.

## Materials and methods

**Focused ion beam system**

We used an FIB/SEM (Helios 5 UX, Thermo Fisher Scientific) with the latest automation software (AutoTEM 5, Thermo Fisher Scientific). A motorized manipulator (EasyLift, Thermo Fisher Scientific) is integrated with the FIB/SEM and controlled via AutoTEM 5 software. Before running AutoTEM 5, we set a chunking position in a bulk, an attaching position in a TEM grid, and FIB parameters. Then, AutoTEM 5 robotically controls all FIB operations of the rough-milling, the lift-out, and the final-thinning process, without the attendance of an operator. The final step of AutoTEM 5 is the 2 kV-FIB polish of a pre-defined region for a pre-defined duration. See the following sections for the FIB parameters used for Si, $SrTiO_3$, and sapphire. Note that the FIB and SEM images in Fig. 1 were manually acquired during the automation process.

**Preparing STEM samples of single-crystal Si**

For Si samples, a protective layer was formed on the surface of single-crystal Si by using FIB-deposition of carbon. Then, a chunk was formed with the dimension of 12 μm by 5.0 μm by 2.4 μm. After lift-out onto a Cu grid, the chunk was thinned by 30kV-FIB. The FIB current was reduced from 24 nA to 26 pA. Then, the sample was polished by 5 kV-FIB with the current of 63 pA. Finally, the sample was polished by 2 kV-FIB with the current of 66 pA. In Fig. 2b, we prepared five Si samples with variable over-tilt angles at 5 kV and 2 kV-FIB as (0.6°, 0.8°), (0.9°, 1.2°), (1.2°, 1.6°), (1.5°, 2.0°), and (3.75°, 5.0°), where (1.2°, 1.6°) corresponds to the Si-optimal recipe.

**Preparing STEM samples of $SrTiO_3$ and sapphire**

First, we sputtered Cr on the surface of single crystals of $SrTiO_3$ and sapphire to avoid charging effect. Then, the bulk crystals were loaded into FIB/SEM chamber. FIB parameters of acceleration voltages and currents are identical with the Si-optimal ones, except for the milling duration. In $SrTiO_3$ and sapphire, the duration in all FIB processes is elongated from the Si-optimal recipe by a factor of three and four, respectively.

**Conventional transmission electron microscopy**

We used a conventional STEM (Talos F200X, Thermo Fisher Scientific) operated at 200 kV for optimizing FIB parameters of Si, $SrTiO_3$, and sapphire.

**Aberration-corrected transmission electron microscopy**

For deep sub-angstrom atomic-level STEM imaging, we used an aberration-corrected STEM (Spectra 300 X-CFEG, Thermo Fisher Scientific) operated at 300 kV. The specification resolution of Spectra 300 is 50 pm at 300 kV. We acquired EDS and EELS mapping by using quadrant EDS detectors (Super-X, Thermo Fisher Scientific) and EELS spectrometer (Continuum ER1065, Gatan) on Spectra 300. Mean free path ($\lambda$) of EELS is estimated as 104 nm (Si), 98 nm ($SrTiO_3$), and 115 nm (sapphire).

We used OptiSTEM+ software (Thermo Fisher Scientific) for acquiring atomic-level STEM images with deep sub-angstrom resolution. The software automatically corrects defocus, 2-fold astigmatism, coma, and 3-fold astigmatism at the region of interest. We note that OptiSTEM+ minimizes artificial deviations of aligning electron optics, providing quantitative reproducibility of deep sub-angstrom resolution to atomic-level STEM imaging. In a post-acquisition processing, drift-corrected frame integration is applied for atomic-level images of both high-angle annular dark-field (HAADF) and EDS mapping. For atomic-level EDS mappings, Radial Wiener filter is also applied.

## Results and discussion

**Robustness of robotic FIB operation**

Figure 1 summarizes the robotic FIB operation of the main process steps, which are rough milling, lift-out, and final thinning. In the beginning, FIB is used to deposit a protective layer (carbon) and cut out a chunk from the single crystal Si (the rough-milling process) (see Fig. 1a). Then an in-situ motorized manipulator is used to lift out the chunk and transfer it to the side position of a TEM grid (the lift-out process) (Figs. 1b, c). Since the lift-out process is performed using automatic image recognition (and not by detecting current flow when the chunk touches the TEM grid), our robotic FIB operation also works on non-conducting samples (see the results from $SrTiO_3$ and sapphire in a later part of this study). In the final-thinning process, FIB milling at 30 kV thins the chunk until the nominal thickness reaches 200 nm (Fig. 1d). Throughout the final-thinning process, the FIB repeatedly corrects translational drift and focus using an alignment marker (the X-shaped pattern in Figs. 1d-f). After thinning at 30 kV, the FIB automatically polishes the lamella at 5 kV and 2 kV (Figs. 1e, f). The nominal sample thicknesses after 5 kV and 2 kV polishing are 120 nm and 60 nm, respectively. The automation software can reliably recognize the alignment marker in FIB images at both 5 kV and 2 kV and thus set the milling regions precisely. The final step in the robotic FIB operation is the 2 kV-FIB polish of a pre-defined region for a pre-defined duration. See also Methods for the FIB parameters used in this study.

Table 1 summarizes the success rate of the robotic FIB experiments for preparing lamellae. In all of our experiments of Si samples (N = 30), the robotic FIB completed all the processes of rough-milling, lift-out, and the final-thinning (including 2 kV-FIB polishing). For $SrTiO_3$ and sapphire, the robotic FIB system also completed all the processes except for one failure in a $SrTiO_3$ sample. The success rates of Si, $SrTiO_3$, and sapphire were 100%, 95%, and 100%, respectively, in our experiments. After completing the robotic FIB operations, all the lamellae were evaluated by STEM and all showed atomic-level resolution. These results confirm the robustness of our robotic FIB operation even when the material is an insulator. Below, we optimize the parameters of robotic FIB operation for single crystal Si.

**Optimizing the final-thinning process for Si**

The quality of an atomic-level STEM image primarily depends on the final-thinning process, as previous studies have highlighted [12-14]. It is generally agreed that 30 kV-FIB damages a lamella from its surface to a depth of the order of 10 nm [13, 33]. Thus, a cleaning step using low-kV FIB (2 kV or below) is important for reducing the surface damage layer to form a high-quality sample suitable for atomic resolution STEM imaging [12-14]. At low kV, blurring of the FIB image can make it difficult to accurately select the milling location (Fig. 1f). This limitation is typically overcome by using sample over-tilt, so that approximate positioning of the low-kV milling region is sufficient (Fig. 2a). As the over-tilt angle becomes large, Ga ions are implanted into the STEM sample. Furthermore, a high over-tilt angle creates a thickness gradient from the top of the sample (protective layer) to the bottom (edge), which limits the thin area within the sample available for high-quality atomic-level STEM imaging. Hence, the over-tilt angle is selected as a compromise between overcoming FIB blurring, degree of Ga implantation, and sample thickness uniformity.

Although the over-tilt angle may affect the quality of the resulting STEM image, a systematic study to determine the optimal over-tilt angle has not been reported previously. To maximize the quality of the sample surface, we study the impact of over-tilt angle on STEM images. We prepared lamellae from single crystal Si with the 2 kV-FIB over-tilt angle ranging from 0.8° to 5.0° (see also Methods). The resulting lamellae were evaluated by conventional STEM (see Methods). To decrease the thickness effect, we acquired atomic-level HAADF images in a region near the top protective layer for each Si lamella (Supplementary Fig. 1). We quantify the degree of sample crystallinity from an atomic-level STEM image by separating the raw image into a crystalline and noise image (see detail in Supplemental Methods and Supplementary Fig. 1). Fig. 2b shows the degree of sample crystallinity of the resulting lamellae as a function of the over-tilt angle used for the 2 kV-FIB polishing step. The degree of crystallinity is highest when using an over-tilt angle of 1.6°.

The optimal over-tilt angle in our study is much lower than that of previous reports (5° ~ 7° at 2 kV) [14]. We note that the FIB/SEM system type used is different from previous studies. Our latest FIB/SEM should have a sharper low-kV FIB beam profile and might shift the trade-off between blurring and ion damage towards a lower over-tilt angle. The targeted sample shape is also different. We aimed to achieve samples with uniform thickness in the lamella plane whereas the previous studies aimed to create wedge-shaped lamellae thinnest at the bottom edge. In other words, the optimal over-tilt angle will depend on the FIB/SEM tool used and the target profile of the lamellae (parallel-sided or wedge-shaped). In our study, we established the optimal over-tilt angle for single crystal Si, at least for our evaluation criteria.

After optimizing the over-tilt angle at 1.6°, the thickness of the Si lamella was fine-tuned by adjusting the milling position offset of the 2 kV-FIB polish (see Supplementary Fig. 2). Next, we characterize the quality and reproducibility of the optimized Si lamellae.

**Characterizing optimized Si lamellae by aberration-corrected STEM**

We prepared three Si lamellae by robotic FIB with the optimal settings for silicon (Fig. 3a). Fig. 3b shows thickness maps of the three lamellae acquired using electron energy loss spectroscopy (EELS). Measuring vertically from the top protective layer towards a bottom edge, the sample thickness increases rapidly from 60 nm to 80 nm and then gradually decreases to ∼ 40 nm. The taper angle between the faces is ∼ 1° (Fig. 3c). In the region near the bottom edge, the thickness across the window width is almost constant at 40 nm ~ 50 nm (Fig. 3d). This sample thickness is suitable for high-quality atomic-level STEM imaging. And the thickness variation among the three samples was only 10 nm (Figs. 3c, d). We emphasize that this profile and thickness reproducibility is a benefit of the robotic procedure and that such reproducibility is important for direct comparison between multiple samples in data-driven research.

We evaluate the suitability of the Si lamellae for aberration-corrected STEM imaging (see Methods). Fig. 4a shows atomic-level HAADF-STEM images from regions near the sample edge. The Si dumbbell structure is visible in all three images. Fast Fourier transforms (FFTs) confirm 55 pm spatial resolution in all the lamellae (Fig. 4b). We also acquired high-resolution HAADF-STEM images in regions near the top protective layers, yielding spatial resolution of 55 pm (Supplementary Fig. 3). We note that the guaranteed resolution of our STEM is 50 pm when using 300 kV electrons. So, our robotic FIB system realizes repeatable fabrication of high-quality lamellae that yield deep sub-angstrom resolution, close to the specified resolution of the latest aberration-corrected STEM instruments.

**Robotic fabrication of high-quality lamellae from SrTiO$_3$ and sapphire**

We next evaluated robotic fabrication using other materials. As shown with Si, the over-tilt angle used for 2 kV cleaning affects the quality of STEM samples. Whereas the final sample thickness depends strongly on the milling rate of each material and thus the procedure needs to be adjusted, validating that the 1.6° over-tilt angle is an effective setting for a range of other materials is also of great importance. This validation matters especially for hard materials, such as SrTiO$_3$, sapphire, diamond, and steel, which are used in various functional materials [16, 18, 21, 22]. We thus tested the robotic sample fabrication on SrTiO$_3$ and sapphire, keeping the over-tilt angle for 2 kV-FIB unchanged at 1.6°.

We modified the Si-optimized procedure for SrTiO$_3$ and sapphire as follows: First, as both materials are insulators (unlike silicon). To suppress charging effects during FIB imaging and milling, we sputtered chromium on the surface of the bulk samples before loading them into the FIB/SEM chamber. Second, to adjust for the different milling rates of these materials, we performed milling rate tests on bulk crystals using 30 kV FIB (Supplementary Fig. 4). Counting the milling rate of Si as 1, the milling rates of SrTiO$_3$ and sapphire are slower at 0.5 and 0.3, respectively. So, we scaled all durations of FIB processes as 3 times and 4 times longer than ones of Si-optimal conditions for SrTiO$_3$ and sapphire, respectively. With this adjusted recipe, our robotic FIB system showed a high success rate of fabricating lamellae from both SrTiO$_3$ and sapphire (Table 1). The robotically fabricated lamellae were evaluated by STEM imaging and all showed atomic resolution. And third, we tuned the sample thickness by adjusting the FIB milling offset, while keeping the over-tilt angle (of 2 kV-FIB milling) constant at 1.6°.

We characterized the resulting lamellae by aberration-corrected STEM. The profile and thickness of both the SrTiO$_3$ and sapphire samples were similar to that of Si (see Fig. 5). All lamellae have a thickness of ∼ 50 nm near the bottom edge. Fig. 6 summarizes the aberration-corrected STEM results of the SrTiO$_3$ lamella. A high contrast HAADF-STEM image in Fig. 6b clearly shows the position of the Sr and Ti atomic columns, and with spatial resolution of 67 pm (Fig. 6c). Energy-dispersive X-ray spectroscopy (EDS) mapping of the SrTiO$_3$ sample also clearly shows the atomic columns of Sr and Ti (Fig. 6d). The sapphire lamella also yields 64 pm image resolution in aberration-corrected STEM (Fig. 7). These results on SrTiO$_3$ and sapphire suggest that the over-tilt angle of 1.6° in the Si-optimal recipe is also suitable for a wide range of hard materials when aberration-corrected STEM imaging is performed.

**Throughput of robotic fabrication of high-quality lamellae**

Regarding the throughput, the robotic fabrication of each lamella took approximately 50 minutes (Si), 60 minutes (SrTiO$_3$), and 80 minutes (sapphire) in our study (see Table 1). Moreover, the automation software can run sequential fabrications of multiple lamellae without the attendance of an operator. The robotic FIB system may increase the total throughput of fabricating high-quality lamellae several times relative to that of the current manual FIB procedure. So, our method will help resolve a current bottleneck of aberration-corrected STEM experiments.

We especially expect that the robotic FIB will play key roles when the aberration-corrected STEM analysis is performed in a study of combinatorial chemistry. A combinatorial method can synthesize a bulk material that has compositional gradients, where a position in the bulk corresponds to its composition [11, 23-26]. In recent studies of combinatorial chemistry, aberration-corrected STEM is sometimes used but the number of lamellae (i.e., the number of studied compositions) is very limited [25, 26]. If the robotic FIB is used, preparing high-quality lamellae at more than ten positions (i.e., more than ten compositions) may become realistic, and aberration-corrected STEM analysis will provide atomic-level insights more in a study of combinatorial chemistry.

**Future challenge**

Despite the proven performance in standard single crystals, the robotic FIB system is still incomplete for complex materials including poly-crystalline materials and layered materials deposited on a substrate. In this study, we systematically optimized the Si recipe and adjusted the Si-optimal recipe to $SrTiO_3$ and sapphire by considering the difference in the bulk milling rate of each material. A similar adaption may be widely effective to a new material. But achieving an optimal final lamella thickness still requires fine-tuning of the milling offset parameters in the final polishing process.

How can we make the robotic FIB smarter and more autonomous in the future? An important challenge of the robotic FIB is to estimate the lamella thickness from information in the monitoring SEM images and use this to automatically determine the termination point of the final FIB polishing. We note that ref [34] reports quantitative measurement of Si lamella thickness using SEM imaging. Closed-loop feedback between FIB polishing and SEM imaging would very likely improve the repeatability of sample thickness and reduce the time needed to optimize the robotic FIB recipe.

## Conclusions

This study demonstrates the automatic fabrication of high-quality lamellae from a wide range of materials using a robotic FIB system. The resulting lamellae of Si, $SrTiO_3$, and sapphire had a thickness of ~ 50 nm near the bottom edge. Aberration-corrected STEM imaging of the lamellae showed deep sub-angstrom spatial resolution of 55 pm (Si), 67 pm ($SrTiO_3$), and 64 pm (sapphire). These results confirm that the quality of robotically-prepared lamellae can meet the requirements of aberration-corrected STEM. Moreover, the repeatability of lamella shape and thickness is a benefit of the robotic FIB operation and plays key roles when STEM results of multiple samples are statistically compared. So, our method will help resolve the current bottleneck of the full workflow of aberration-corrected STEM experiments and accelerate the discovery of new materials based on atomic-level investigations.

## Acknowledgments

The authors thank M. Munekane for preparing Si samples and S. Lazar for arranging STEM measurements. H.T. also thanks A. Bright, M. Dutka, and E. Tochigi for comments on the manuscript. T.M. was financially supported by the Ministry of Education, Culture, Sports, Science and Technology (MEXT); Nos 17H06094, 19H00818, and 19H05787.

## Author contributions

H.T. and T.M. conceived the project. H.T. designed the experiments. K.K. prepared samples. N.N. and Y.C. performed STEM measurements. H.T. wrote the manuscript. All authors discussed the results and commented on the manuscript

## Conflict of interests

H.T., N.N., K.K., Y.C., and B.V.L. have a financial interest in Thermo Fisher Scientific.

## Data and materials availability

All data needed to evaluate the conclusions in the paper are present in the paper and the Supplementary Information. Upon a reasonable request, corresponding authors can provide the FIB recipes for Si, SrTiO$_3$, and sapphire. (Contact H.T.)

35. Tochigi, E. *et al.* Dissociation reaction of the 1/3 <1101> edge dislocation in α-$Al_2O_3$. *Journal of Materials Science* **53**, 8049–8058 (2018).

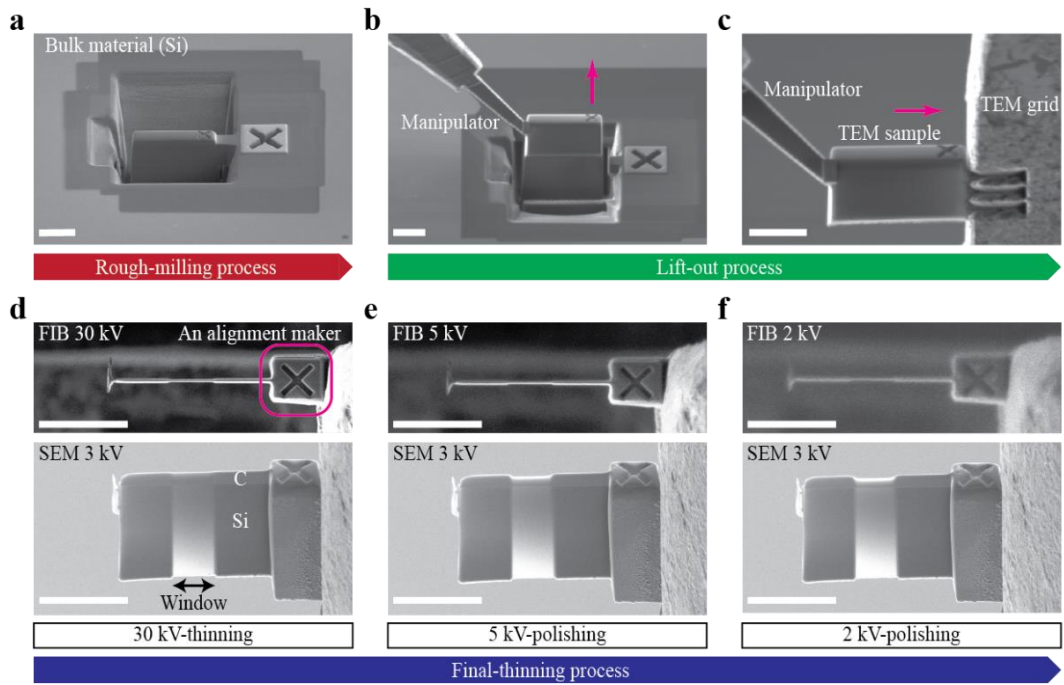

**Fig. 1 Robotic FIB operation to prepare a high-quality lamella. a** Rough-milling process to prepare a chunk in a bulk, **b** The in-situ manipulator lifts out the chunk. **c** The chunk is transferred and fixed to a TEM grid at the position. After the FIB then cuts the tip of the manipulator to release the sample, the final-thinning process starts. **d** FIB and SEM (secondary electron) images after 30 kV thinning. Finally, the FIB thins and polishes a narrower "window" region of approximately 2.4 μm wide. **e** and **f** FIB and SEM images after FIB polishing at 5 kV and 2 kV, respectively. Preparing one STEM sample takes about 1 hour (see Table 1). The scale bars shown in both FIB and SEM images are 5 μm long.

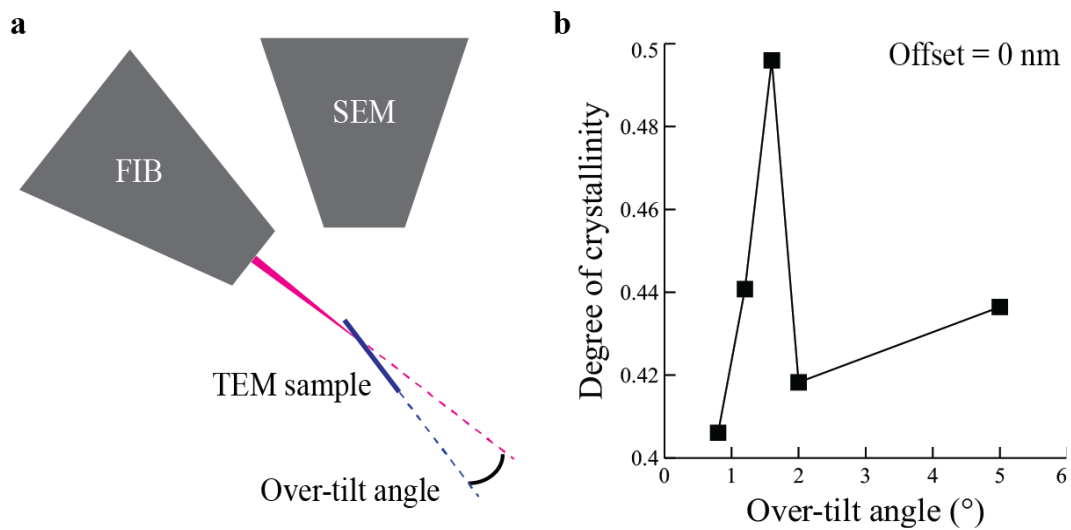

**Fig. 2 Optimizing FIB parameters for Si. a** A side-view diagram of the over-tilt angle, which is the angle between the plane of the TEM sample and the ion beam. **b** A plot of the measured degree of silicon crystallinity as a function of the over-tilt angle used for the 2 kV-FIB polishing step. We prepared five lamellae, using over-tilt angles of 0.8°, 1.2°, 1.6°, 2.0°, and 5.0°, respectively. The degree of crystallinity was calculated from the high-resolution STEM images of each sample (see Methods and Supplementary Fig. 1).

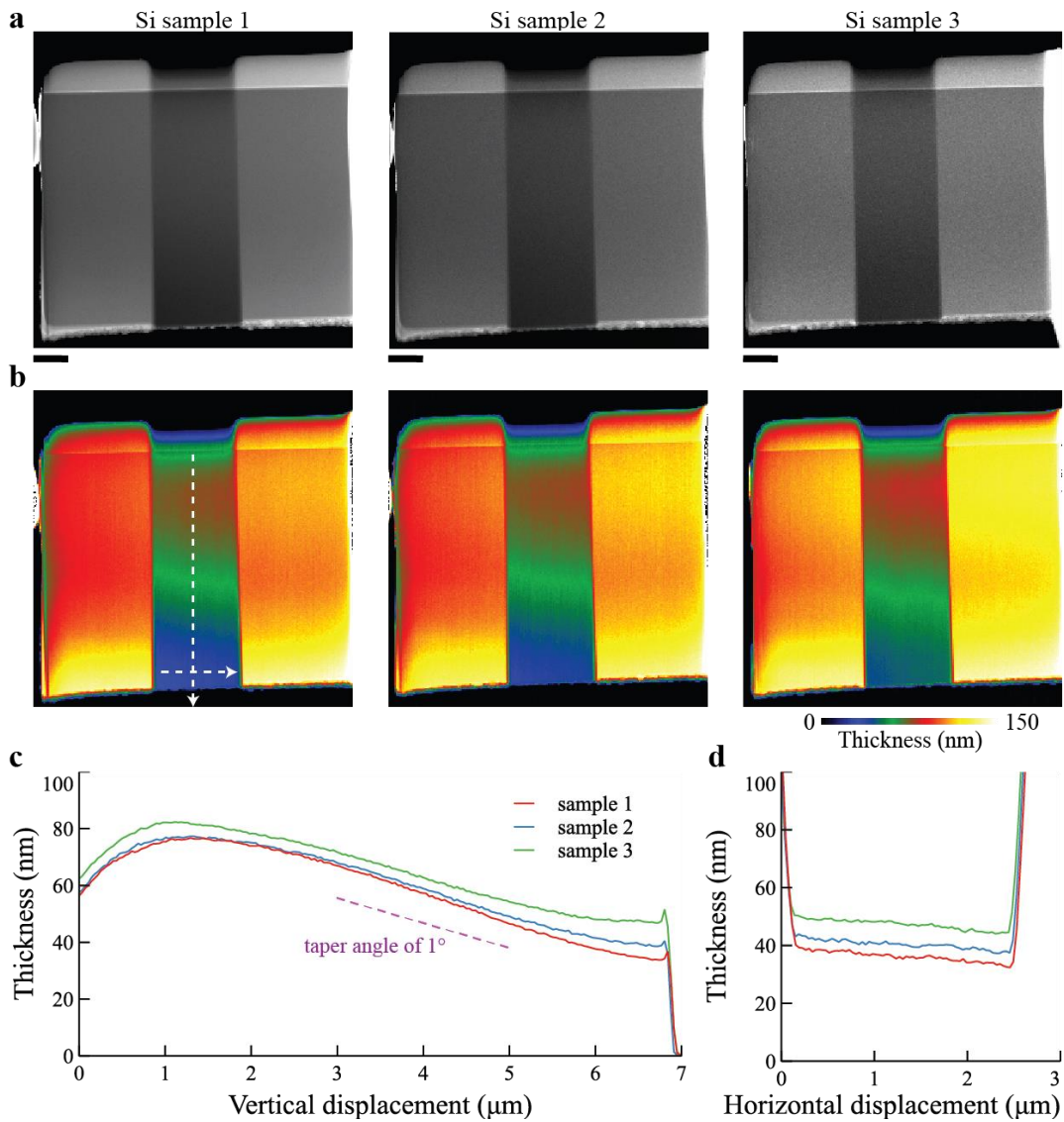

**Fig. 3 Sample shape and thickness reproducibility of robotically prepared Si lamellae. a** HAADF-STEM images of each sample. The three lamellae were fabricated using the Si-optimized settings. **b** Thickness maps generated from EELS spectrum imaging. **c** Thickness profiles in the vertical direction (from the top protective layer to the bottom edge). The dashed line shows a 1.0° taper angle. **d** Thickness profiles horizontally across the thin window near the bottom edge. The scale bar is 1 μm.

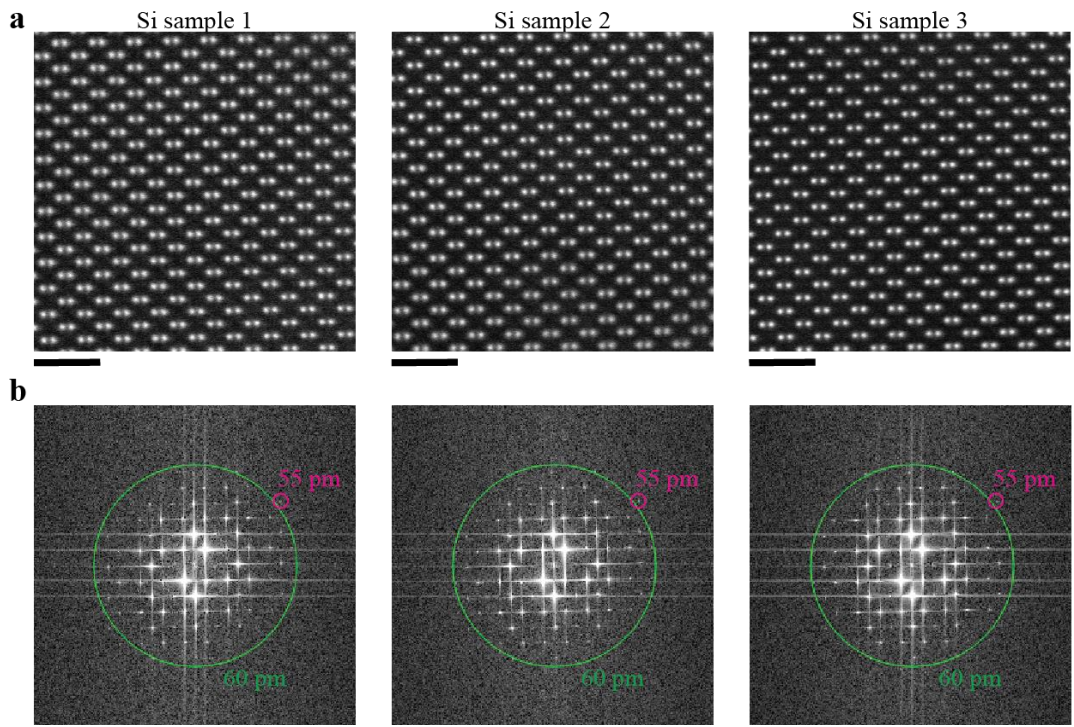

**Fig. 4 Aberration-corrected STEM imaging of the three Si lamellae. a** Atomic-level HAADF-STEM images of each lamella along the [110] direction acquired by aberration-corrected STEM (same samples as Fig. 3). Images were acquired near the thin bottom edge of each sample. Equivalent HAADF-STEM images from near the top protective layer are shown in Supplementary Fig. 3. **b** FFT patterns for each image. Magenta circles indicate FFT reflections at 55 pm resolution. Green circles correspond to 60 pm resolution. The scale bar is 1 nm.

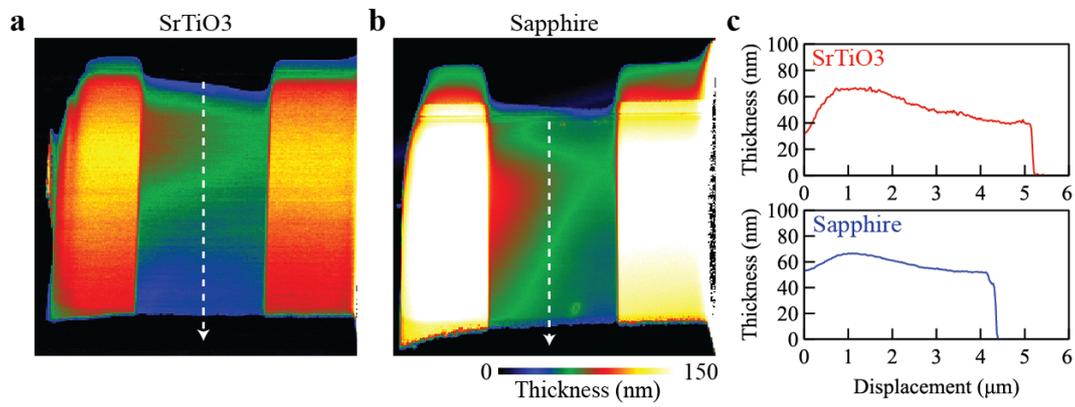

**Fig. 5 Sample thickness of SrTiO₃ and sapphire lamellae prepared by the optimized robotic FIB. a** and **b** Thickness maps from low loss EELS spectrum imaging of SrTiO₃ and sapphire, respectively. **c** Thickness profiles from the top protective layer to the bottom edge.

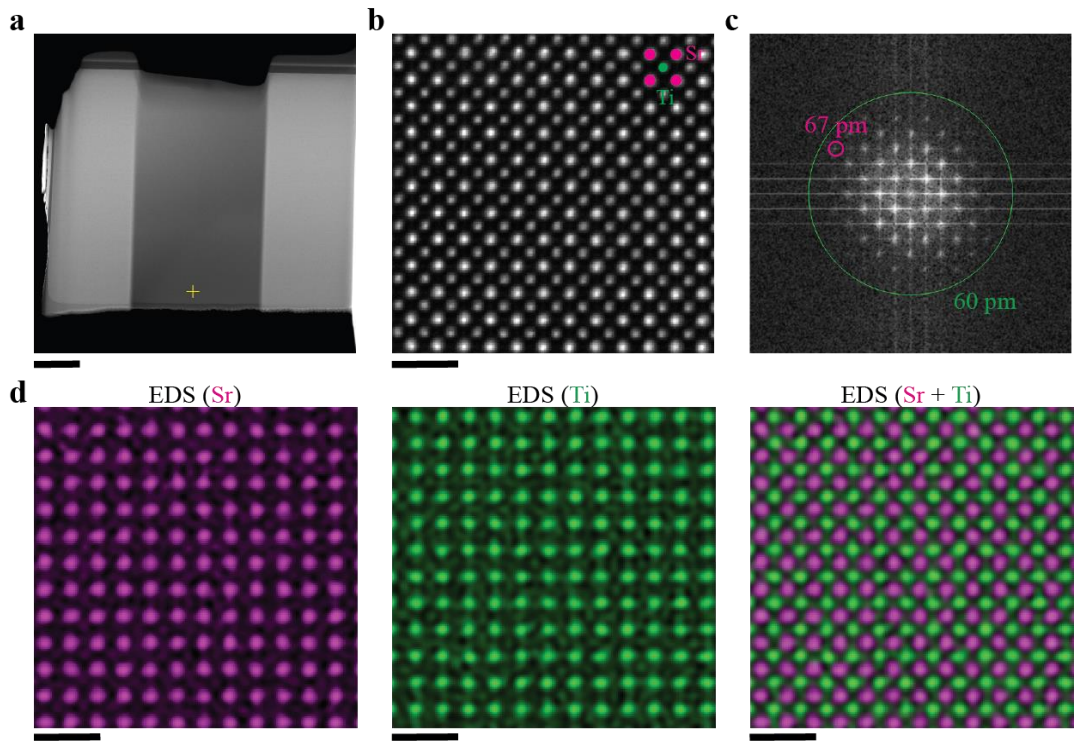

**Fig. 6 Aberration-corrected STEM results of a robotically prepared SrTiO$_3$ lamella. a** A HAADF-STEM image of the entire SrTiO$_3$ lamella. **b** An atomic-level HAADF-STEM image along the [100] direction. The image was acquired near the thin bottom edge of the SrTiO$_3$ lamella (marked + in **a**). **c** FFT pattern of **b**. The magenta circle indicates an FFT reflection at 67 pm resolution. **d** Atomic-level EDS maps for Sr (left), Ti (center), and Sr combined with Ti (right). The scale bar in **a** is 1 μm, that in **b**, and **d** is 1 nm.

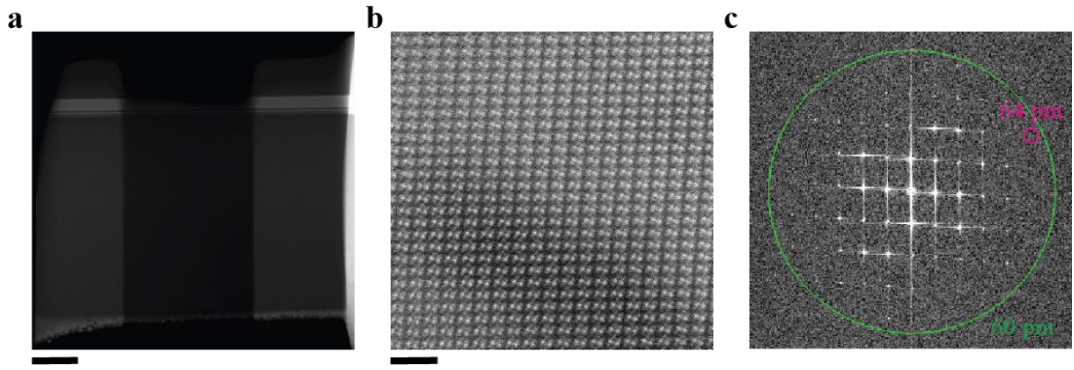

**Fig. 7 Aberration-corrected STEM imaging of a robotically prepared sapphire lamella. a** A HAADF-STEM image of the entire sapphire lamella. **b** An atomic-level HAADF-STEM image near the edge of the sapphire sample. See ref [35] for the atomic model of this material. **c** FFT pattern of **b**. Scale bars in **a** and **b** correspond to 1 μm and 1 nm, respectively.

|  | Si | SrTiO$_3$ | Sapphire |
|---|---|---|---|
| **# of experiments[A]** | 30 | 20 | 2 |
| **# of successful fabrications[B]** | 30 | 19 | 2 |
| **Atomic-level STEM images** | 30 | 19 | 2 |
| **Success rate (B/A)** | 100% | 95% | 100% |
| **Throughput of fabrication (minutes per lamella)** | 50 | 60 | 80 |

**Table 1 Summary of the robotic FIB experiments.** The number of experiments includes preliminary tests of fabricating lamellae to explore FIB parameters. The successful fabrication counts the number of the experiments that completed all the automation processes (including 2 kV-FIB polishing). One fabrication of SrTiO$_3$ lamella failed in the lift-out process. All the lamellae in "successful fabrications" were evaluated by STEM and yielded atomic-level STEM images. Throughput of fabrication is the duration that the automation software spent in rough-milling, lift-out, and the final-thinning processes of each STEM sample. This value is a typical one and excludes the processing time of loading bulk samples and setting the automation software.

# Supplementary Information for

# Robotic fabrication of high-quality lamellae for aberration-corrected transmission electron microscopy

Hideyo Tsurusawa\*†, Nobuto Nakanishi†, Kayoko Kawano,
Yiqiang Chen, Brandon Van Leer, Teruyasu Mizoguchi\*

†These authors contributed equally to this work.

\* Corresponding authors.
Email: Hideyo.Tsurusawa@thermofisher.com, teru@iis.u-tokyo.ac.jp

Supplementary methods.

Supplementary Fig. 1 Image processing for extracting crystalline degree.

Supplementary Fig. 2 Fine-tuning of the thickness of a Si lamella by milling-offset.

Supplementary Fig. 3 Aberration-corrected STEM imaging of Si lamellae near the top protective layer.

Supplementary Fig. 4 Comparing milling rates between Si, $SrTiO_3$, and sapphire.

**Supplementary methods**

**Computing degree of crystallinity from a STEM image**

We quantify the degree of crystallinity from an atomic-level STEM image of single-crystal Si by the following procedure. An atomic-level STEM image, $I_0(x,y)$, was acquired by conventional STEM (see Supplementary Fig. 1b as an example). The original STEM image is transformed in reciprocal space as an original FFT pattern (Supplementary Fig. 1b inset). Then, the original FFT pattern is filtered into a crystalline pattern (including only periodic spots) (Supplementary Fig. 1c inset). The crystalline FFT pattern is inverted into a real-space image as a crystalline image, $I_1(x,y)$ (Supplementary Fig. 1c). Then, a noise image, $I_2(x,y)$, is also extracted as $I_2(x,y) = I_0(x,y) - I_1(x,y)$. Instead of simple averaging of image intensity, we calculate the standard deviation of the image intensity because the baseline of image intensity depends on the amplifier settings of every image acquisition. Hence, crystalline and noise scores, $\sigma_1$ and $\sigma_2$, are defined as the standard deviation of intensity in the crystalline and noise image, respectively: $\sigma_1 = \sqrt{\langle I_1^2(x,y) - \langle I_1(x,y)\rangle^2\rangle}$ and $\sigma_2 = \sqrt{\langle I_2^2(x,y) - \langle I_2(x,y)\rangle^2\rangle}$, where $\langle\ \rangle$ averages over $(x,y)$ in an image. Finally, the degree of crystallinity is computed as $\sigma_1/(\sigma_1 + \sigma_2)$. This order parameter can range from 0 (complete amorphous) to 1 (complete crystal).

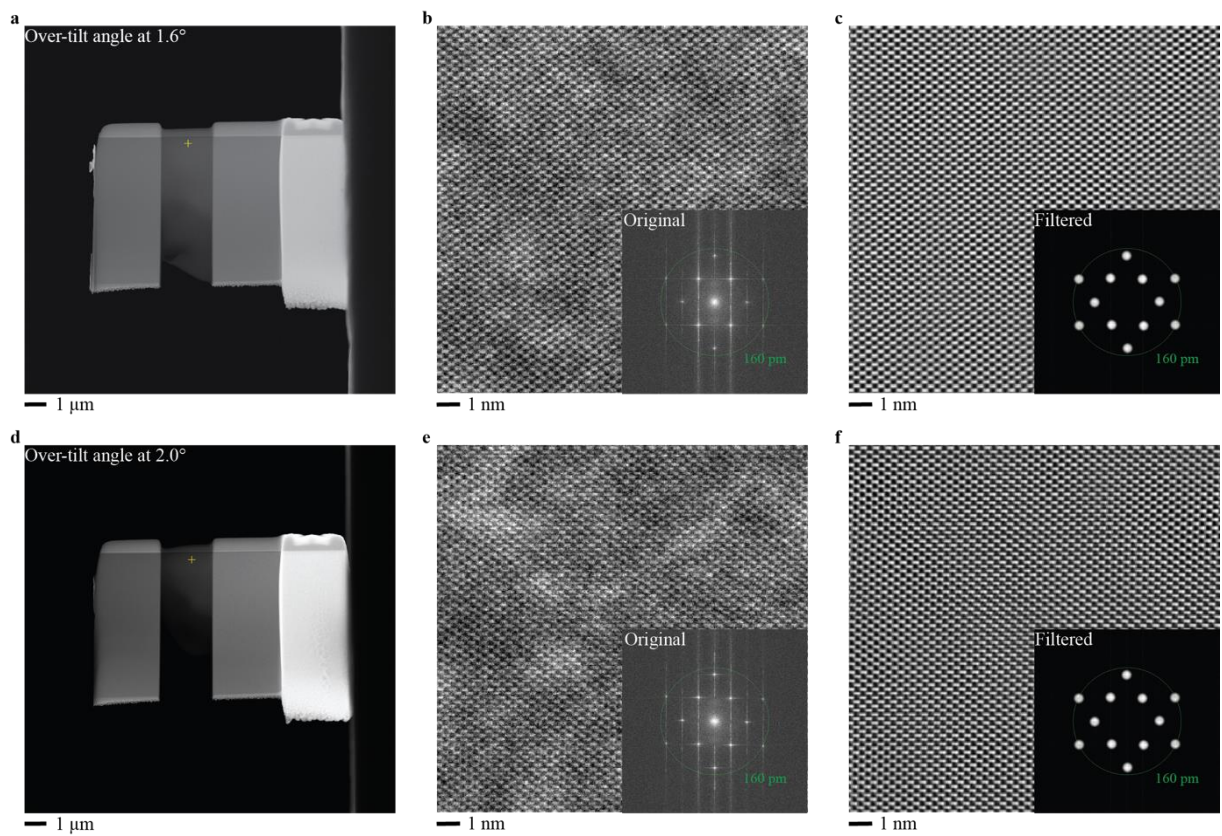

**Supplementary Fig. 1 Image processing for extracting crystalline degree. a-c** STEM results from a Si lamella, in which over-tilt angle at 2 kV-FIB is set as 1.6°. **a** A low-magnification HAADF-STEM image. **b** An atomic-level HAADF image was acquired in a region near the top protective layer (+ mark in **a**) along the [110] direction. Inset is the original FFT pattern of the original HAADF image. **c** A crystalline image from **b**. Original FFT pattern is masked to pass only periodic pattern of Si lattice. Inset shows the masked crystalline FFT pattern. Then, the crystalline image is computed by invert-FFT from the crystalline FFT pattern. **d-f** STEM results from another Si lamella, in which over-tilt angle at 2 kV-FIB is set as 2.0°. All images were acquired by a conventional TEM, operated at 200 kV. To improve statistics, we acquired STEM images with the size of 47.7 nm by 47.7 nm. Images in **b**, **c**, **e**, and **f** are cropped from the raw images.

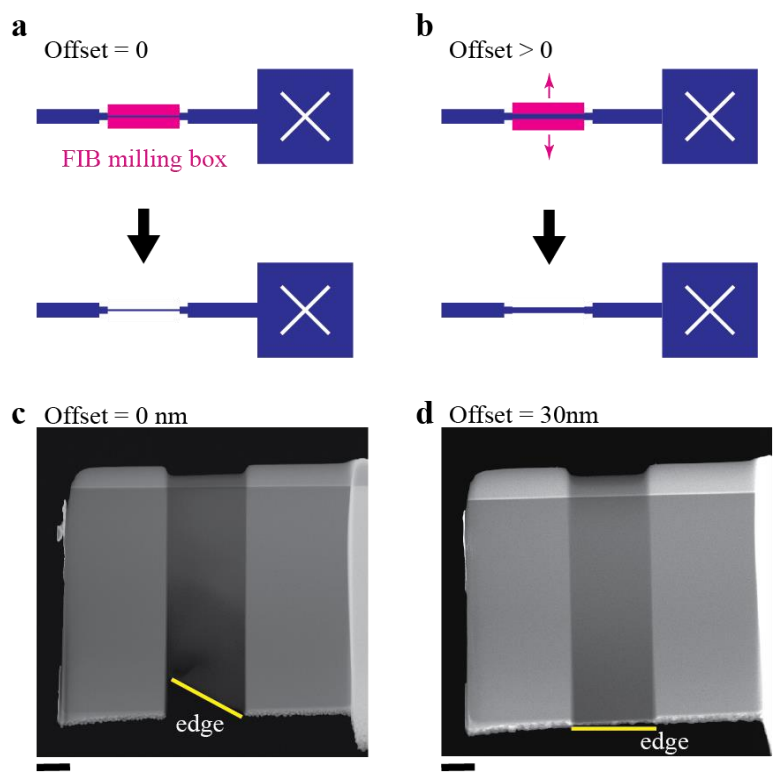

**Supplementary Fig. 2 Fine-tuning of the thickness of a Si lamella by milling-offset. a** Schematics of top-down FIB showing the effect of changing the milling offset (both sides) in the recipe to control the sample thickness. The automation software sets the position of the FIB milling box based on the target thickness. The milling offset shifts the final FIB milling boxes away from the sample center. **b** When an offset is added (offset > 0), the resulting final thickness increases. **c, d** HAADF-STEM images of Si samples prepared using a fixed 1.6° over-tilt angle and using a 2 kV-FIB milling offset of 0 nm (in **c**) and 30 nm (in **d**). In the case of zero offset, the Si sample was over-polished and the target area was lost (see the bottom shape in **c**). With an offset of 30 nm, the bottom edge of the sample is very thin but is still present (see the bottom edge in **d**). Hence, we tune the thickness of a Si lamella by applying the milling offset of 30 nm at the final 2 kV-FIB process (where the over-tilt angle is set at 1.6°). Scale bars are 1 μm.

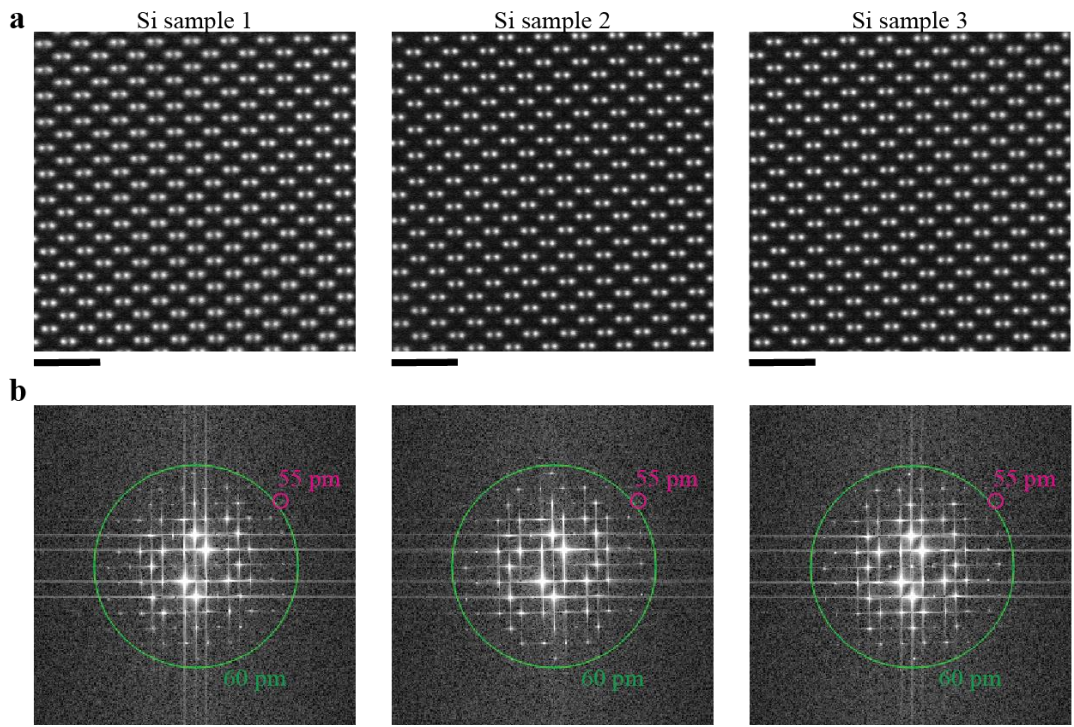

**Supplementary Fig. 3 Aberration-corrected STEM imaging of Si lamellae near the top protective layer. a** Atomic-level HAADF-STEM images of Si lamellae along the [110] direction. The three lamellae are the same as in Figs. 4 and 5. Each atomic-level HAADF-STEM image was acquired in regions near the top protective layer. **b** FFT patterns of **a**. Magenta circles are eye-guides of diffraction spots of 55 pm resolution. Green circles correspond to 60 pm resolution. The scale bar is 1 nm.

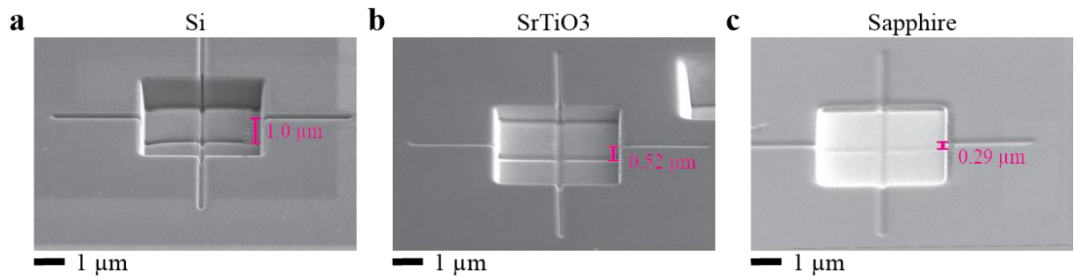

**Supplementary Fig. 4 Comparing milling rates between Si, SrTiO₃, and sapphire.** FIB at 30 kV milled bulks of Si, SrTiO₃, and sapphire, where FIB current and milling duration are all the same. Then, SEM measures the milling depths of Si, SrTiO₃, and sapphire as 1.0 μm, 0.52 μm, and 0.29 μm, respectively.